\documentclass[runningheads]{llncs}
\usepackage{float}
\usepackage{caption}
\usepackage{subcaption}
\usepackage{pgf-pie}

\usepackage{smartdiagram}
\usesmartdiagramlibrary{additions}
\usepackage{graphicx}
\usepackage[colorlinks,bookmarksopen,bookmarksnumbered,citecolor=red,urlcolor=red]{hyperref}

\makeatletter
\def\BState{\State\hskip-\ALG@thistlm}
\makeatother

\begin{document}

\title{\textit{BACKUS}: Comprehensive High-Performance Research Software Engineering Approach for Simulations in Supercomputing Systems}

\titlerunning{\textit{BACKUS}}

\author{Matan Rusanovsky\inst{1,2}
\and Re'em Harel\inst{2,3}
\and Lee-or Alon\inst{2,4}
\and Idan Mosseri\inst{1,6}
\and Harel Levin\inst{5,6}
\and Gal Oren\inst{1,6}}

\authorrunning{Rusanovsky, Harel, Alon, Mosseri, Levin, and Oren}

\institute{
Department of Computer Science, Ben-Gurion University of the Negev, P.O.B. 653, Be'er Sheva, Israel
\and Israel Atomic Energy Commission, P.O.B. 7061, Tel Aviv 61070, Israel
\and Department of Physics, Bar-Ilan University, IL52900, Ramat-Gan, Israel
\and Department of Computer Science, Bar-Ilan University, IL52900, Ramat-Gan, Israel
\and Department of Mathematics and Computer Science, The Open University of Israel, P.O.B. 808, Ra'anana, Israel
\and Department of Physics, Nuclear Research Center-Negev, P.O.B. 9001, Be'er-Sheva, Israel\\
\email{matanru@post.bgu.ac.il, reemharel22@gmail.com, lee-or@mail.com, idanmos@post.bgu.ac.il, harellevin@gmail.com, orenw@post.bgu.ac.il}}

\maketitle

\begin{abstract}
High-Performance Computing (HPC) platforms enable scientific software to achieve breakthroughs in many research fields such as physics, biology, and chemistry, by employing Research Software Engineering (RSE) techniques. These include 1) novel parallelism paradigms such as Shared Memory Parallelism (with e.g. OpenMP 4.5); Distributed Memory Parallelism (with e.g. MPI 4); Hybrid Parallelism which combines them; and Heterogeneous Parallelism (for CPUs, co-processors and accelerators), 2) introducing advanced Software Engineering concepts such as Object Oriented Parallel Programming (OOPP); Parallel Unit testing; Parallel I/O Formats; Hybrid Parallel Visualization; and 3) Selecting the Best Practices in other necessary areas such as User Interface; Automatic Documentation; Version Control and Project Management. In this work we present \textit{BACKUS}: Comprehensive High-Performance Research Software Engineering Approach for Simulations in Supercomputing Systems, which we found to fit best for long-lived parallel scientific codes.
\end{abstract}

\keywords{Research Software Engineering, High-Performance Computing, Supercomputers, Simulations, Scientific Computing}

\section{Introduction}
\subsection{Simulations in Supercomputing Systems}
In computational scientific computing, studying complex scientific problems requires advanced computing capabilities. This field exists since the 1960s \cite{nonweiler1984computational}, and has been growing rapidly since distributed and parallel computing methods were introduced to the scientific community in the 1990s \cite{armstrong1999toward}. As of today, High Performance Computing (HPC) platforms are being extensively  used in many fields of research, such as physics, biology and chemistry, in order to solve a system of equations. This system is solved using a time and space discretization over the physical domain of the problem \cite{anderson1995computational}. A common parallel approach is to divide the domain to different computing nodes, where each node solves a part of the domain. This naive algorithm may be used in shared memory as well as disturbed memory architectures \cite{babuska2012modeling}. In contrast to “textbook” computer codes, scientific codes are usually hold a substantial number of physical arrays, which are being shared and used by the code routines \cite{harris1964numerical}.
\subsection{Changes in High-Performance Computing Architectures}
Over the first 25 years of high-performance computing, growth of computing power was achieved almost solely based on the single-thread exponential performance increase \cite{KarlRupp1}. Many simulations codes were introduced with the relatively easy to implement Massage Passing Interface \cite{MPIForum} (MPI) which allowed - since its release in 1994 - to distribute computations on distribute-memory clusters and supercomputers, and therefore became the de facto standard for parallel computations. As the codes scaled well over those years, no major changes were requested from developers and hardware architects alike. 

\begin{figure}[H]
  \centering
  \includegraphics[width=\textwidth]{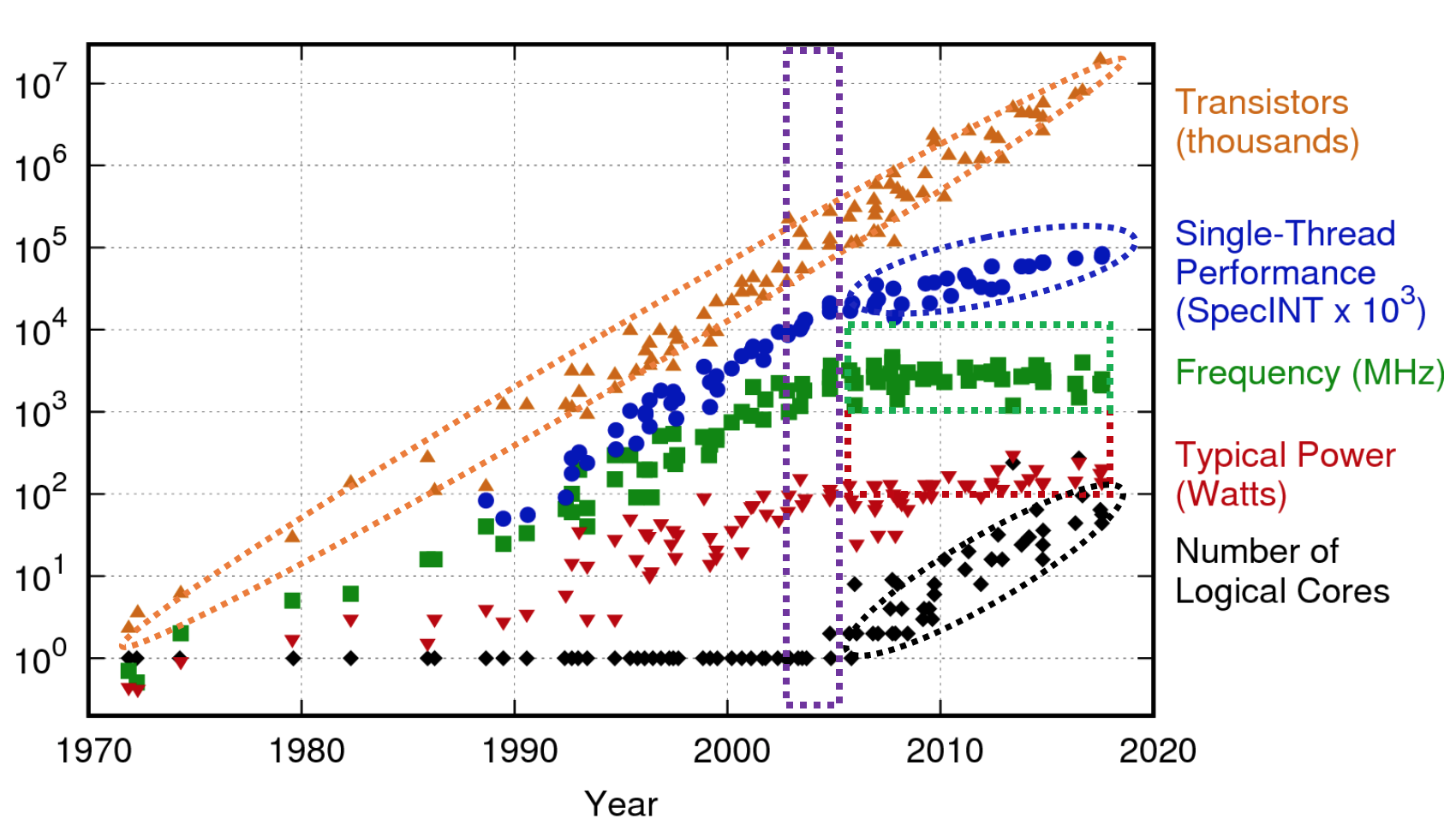}
  \caption{42 Years of Microprocessor Trend Data \cite{KarlRupp1}. Orange: Moore's Law trend; Purpule: Dennard scaling breakdown; Green \& Red: Immidiate implications of Dennard scaling breakdown; Blue: Slowdown of ST increase in performances; Black: The age of increase parallelisem.}
  \label{figure1}
\end{figure}

  \begin{figure}[H]
  \centering
  \begin{subfigure}[b]{\linewidth}
  \centering
  \includegraphics[width=0.8\textwidth]{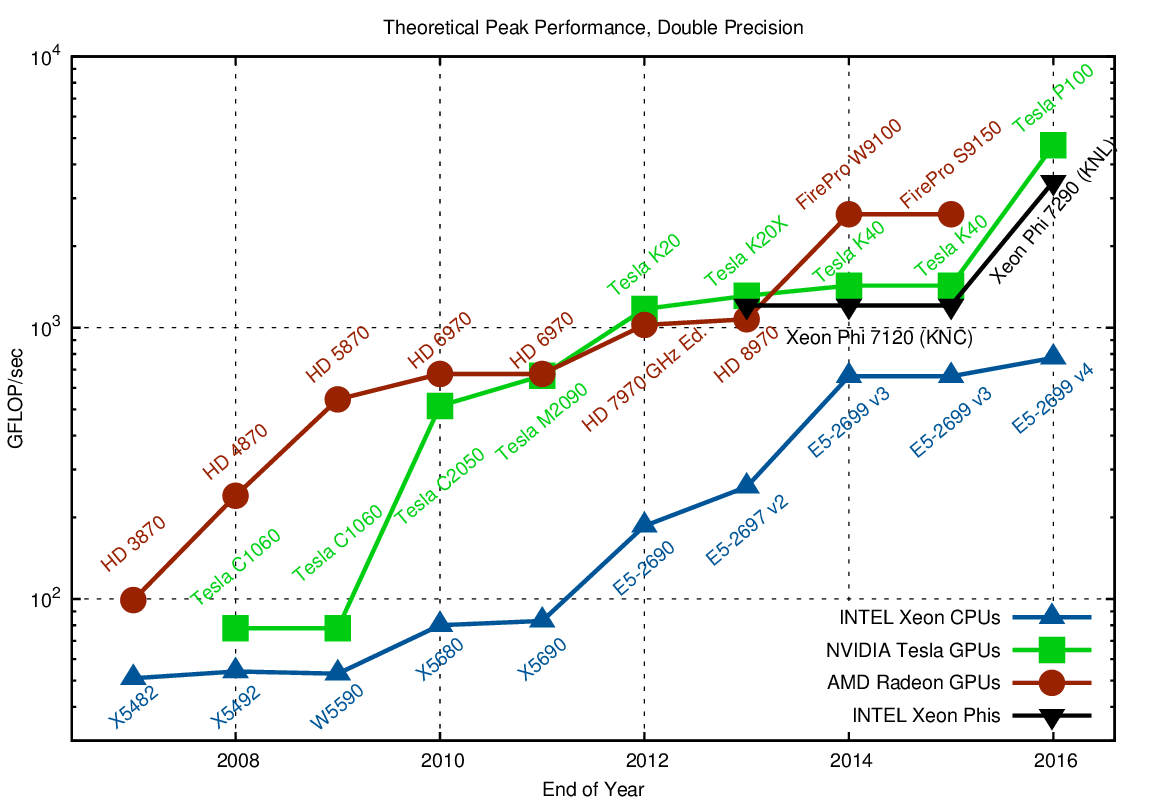}
  \caption{Comparison of theoretical peak GFLOP/sec in double precision.}\label{fig:5}
  \end{subfigure}
  \begin{subfigure}[b]{\linewidth}
  \centering
  \includegraphics[width=0.8\textwidth]{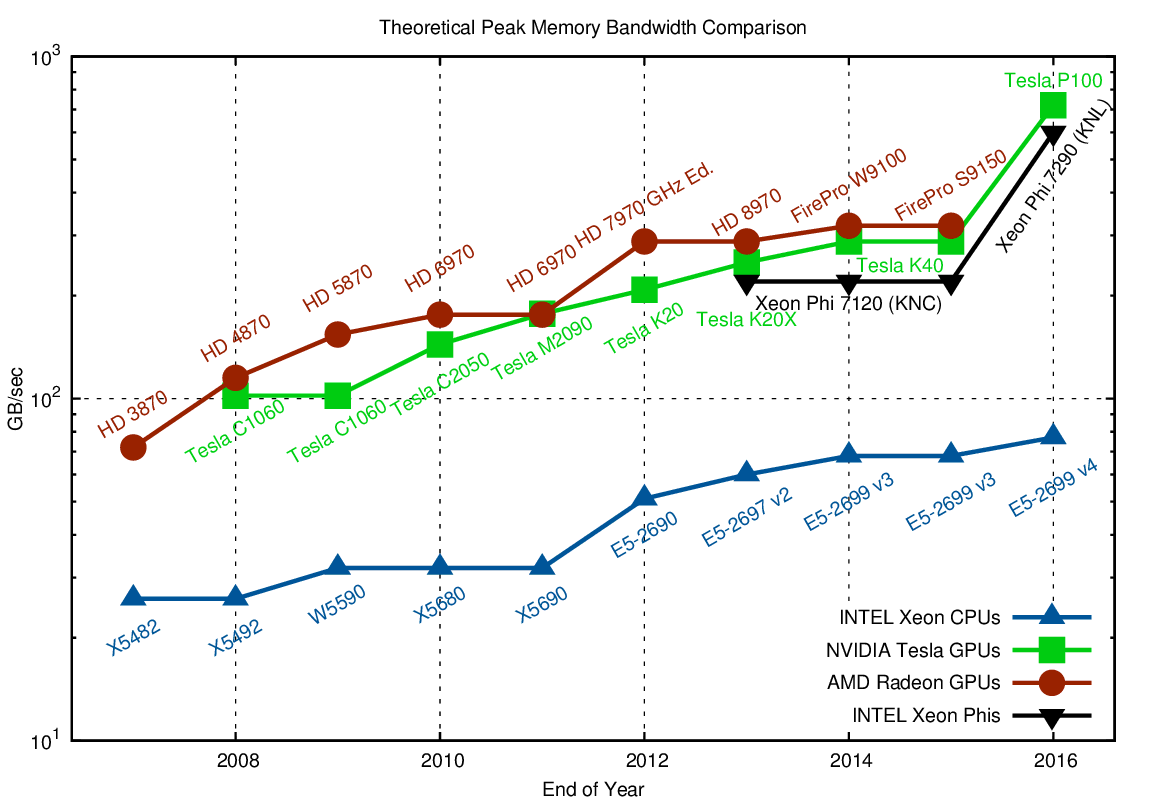}
  \caption{Comparison of theoretical peak memory bandwidth.}\label{fig:4}
  \end{subfigure}
  \caption{CPU, GPU and MIC Hardware Characteristics over Time. \cite{KarlRupp1} \cite{KarlRupp2}}
  \label{figure2}
\end{figure}

However, although Moore's law \cite{moore1965cramming} (still) did not break down, Dennard scaling \cite{dennard1974design} - also known as MOSFET scaling - did. Dennard predicted in 1974 that the power density of transistors will keep to be constant, even as their dimensions will continue to decrease, which in turn will increase clock frequencies constantly and rapidly in order to push performances to the limit, all without raising power consumption. Yet, this scaling did not endure forever, and in 2005 stopped (as demonstrated in Figure \ref{figure1}), resulted in an unfamiliar situation in which Moore's law keep standing, but the yielded performances are no longer dramatic as they use to be \cite{bohr200730}. Therefore, industry started to focus on new hardware architectures - and consecutively code paradigms - to keep trends on track. The most important shift was the introduction of extensive parallelism in the form of increasing the amount of logical cores in CPUs, and the transition to accelerators and co-processors which operate at lower frequencies than CPUs but posses orders of magnitude more cores than the latter. This shift allows most operations to operate in a vectorized fashion, thus increasing performances even further \cite{esmaeilzadeh2011dark}. All of these advancements in turn created a new era in which heterogeneous computing - a mix of several hardware architectures - is almost the only way to keep pushing performances. Figure \ref{figure2} shows a comparison between those architectures in terms of peak performances and memory bandwidth. However, there is no free lunch, and in order to crop the potential performances, major changes have to be made to the code structure.


\subsection{Changes in High-Performance RSE Approaches}
In accordance to the changes in HPC architectures, there was an immanent need to change code structure fundamentally in order to exploit the full benefits of the hardware. Among those changes include support of Object-Oriented Parallel Programming; Using shared memory with MPI and binding MPI to support those underlining directives (Hybrid Programming); Detailed attention to cache-coherent Non-Uniform Memory Access topology (cc-NUMA) using Thread Affinity; Utilization of Single-Instruction Multiple Data (SIMD) capabilities - both explicitly and implicitly; Parallel I/O which target the Distributed File System management and physical capabilities; Enhanced fault-tolerance as the amount of cores and length of computation grows; and detailed acceleration of parts of the code using data mappings and synchronous/asynchronous offloads/uploads. 

In parallel to those changes, numerous changes were introduced to the world of Research Software Engineering (RSE) in general, resulting in a major needed change of legacy codes standards in the aspects of Development Process and Environment; Testing, Validation and Verification (V\&V); Project Management and Documentation; and User Interface and Experience (UI/UX), which became crucial as those codes became more and more complex \cite{feathers2004working}. These changes also were driven from the dramatic changes in personnel selection and the marketplace that do not assume tenure workers who are familiar with the code, meaning that the need to create a set of RSE standards in order to constantly handshake the code development became crucial. Moreover, the Open-Source reality, that assumes code development by other contributors that are not part of the core code developers, forced long-lived codes to be as clear and as dynamic as possible in order to keep its relevance and usability.

As a result, there was a clear need to form a comprehensive high-performance RSE approach for simulations in modern supercomputing systems, which treats \textit{all} of the aspects of a new simulation development in the current reality - from the most minor to the most significant ones. We call this approach \textit{BACKUS} (Figure \ref{figure3}), after the late pioneer \textit{John Backus}, who turned Fortran - the most wildly used language in simulations coding - to the first high-level programming language to be put to broad use \cite{BACKUS1957fortran}.

We limit the discussion in \textit{BACKUS} regarding specific compilers, debuggers and profilers as we found those strongly coupled with the selected hardware's vendor, and thus not representing a \textit{general} perspective on high-performance RSE. Moreover, one of the most important guidelines for the following selected methods in \textit{BACKUS} is the degree of freedom the tools have from vendors as well as hardware.

\section{\textit{BACKUS}: Comprehensive High-Performance RSE Approach for Simulations in Supercomputing Systems}

In order to create a long-lasting scientific production code with top performances on supercomputers, we concluded that there must be a complete change of heart in the way legacy scientific codes are written, developed and maintained. This approach derives from two main reasons:
\begin{itemize}
\item An introduction of a sophisticated parallelism is almost impossible when the code is not structured well, i.e., code which is written hastily and hard to understand; Based on ad-hoc design instead of long-lasting one; Code which is hard to extend and make changes to, especially when considering new parallelism paradigms; Difficult to optimize for new computer architectures; Impossible to test without thoroughly be familiar with the code behavior and so forth \cite{ferenbaugh2018bringing}.
\item This introduction of sophisticated parallelism paradigms is mostly done today by HPC scientists instead of natural scientists as used to be a decade ago, mostly as the needed complexity of the code to achieve top performances increased dramatically, and therefore the inability of the first ones to control it will yield with the inability to introduce it with any kind of change or improvement in the future.
\end{itemize}
Therefore, a modern scientific code needs to be understandable, maintainable, extensible, well-tested, well-documented, portable to modern architectures and most importantly - to be constructed in a way which will allow it to take advantage of the optimal parallelism methods \cite{feathers2004working}. As a consequence, we formed the following suite, which we found to be as optimal as possible to the creation of a new long-lasting parallel scientific code:

\begin{figure}[H]
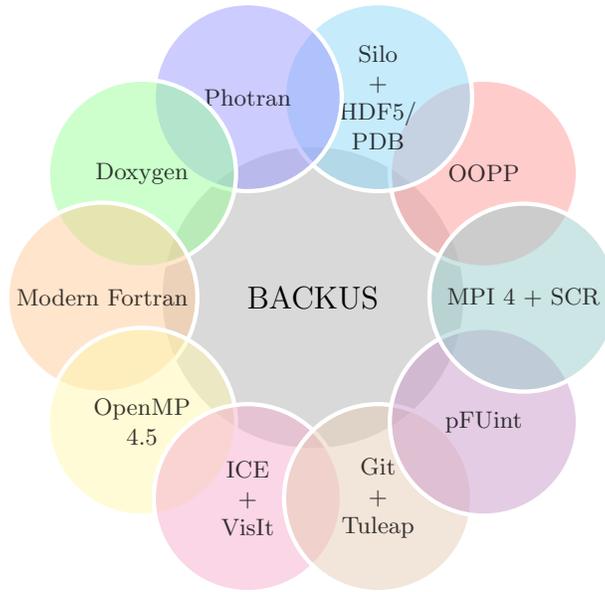

  \centering

\smartdiagramset{circular distance=4cm,
font=\small,
text width=1cm,
module minimum width=1.5cm,
module minimum height=1.5cm,
arrow tip=to}
\smartdiagram[bubble diagram]{BACKUS, OOPP, Silo\\ + \\HDF5/\\PDB, Photran, Doxygen,  Modern Fortran, OpenMP \\4.5, ICE\\ + \\VisIt, Git\\ + \\Tuleap,  pFUint,  MPI 4 + SCR}

  \caption{\textit{BACKUS}: Comprehensive High-Performance Research Software Engineering Approach for Simulations in Supercomputing Systems.}
  \label{figure3}
\end{figure}

  \subsection{ \textbf{Programming Language:} \textit{Modern Fortran}}

  \textit{Modern Fortran} (“\textbf{For}mula \textbf{Tran}slation”) has been developed by John Backus and coworkers in 1957 for IBM, and it has been updated every several years since then. Fortran is excellent for array/matrix handling and calculations, hence it is considered to be the undisputed leader for scientific computations \cite{moreira1998comparison}. This derives from the fact that scientific calculations are mostly numerical calculations, which substantially based on elementary arithmetic, in which Fortran exceles in performances. Figure \ref{fig:5} presents a comparison between different languages on core operations that are extensively used in numerical and HPC applications, and exemplifies the factors and even the order of magnitudes of performance differences between them in comparison to Fortran (in purpule).  
  
  Fortran 2003 introduced object-oriented programming support, thus enabled to design systems through modern design concepts \cite{hanson2013numerical} \cite{gorelik2004object}, while still benefiting from the efficiency of the language. For example, Haveraaen et al. \cite{haveraaen2015high} presented object-oriented design patterns for HPC applications that achieved high performance scalability even at the presence of tens of thousands of cores execution. 
  
  Another important point is that Fortran was proven to be the best programming language in the arena of scientific computing over the last several decades, hence lots of legacy codes were already written in Fortran \cite{loh2010ideal}. Modern Fortran supports codes programmed in older versions of Fortran, and it even supports interoperability with C, thus still making it the most suited \textit{modern} language for scientific calculations. 
  
    \begin{figure}[H]
  \centering
  \begin{subfigure}[b]{\linewidth}
  \centering
  \includegraphics[width=7cm,height=7cm]{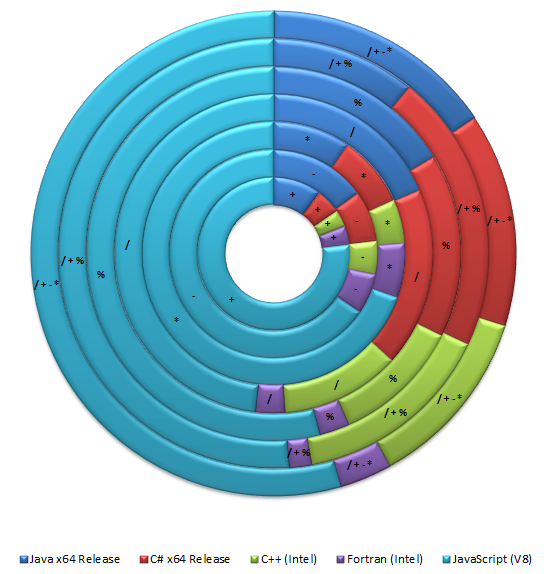}
  \caption{Qualitative comparison of the languages performance on the core operations
  \cite{perfor}.}\label{fig:5}
  \end{subfigure}
  \begin{subfigure}[b]{\linewidth}
  \centering
  \includegraphics[width=0.8\linewidth]{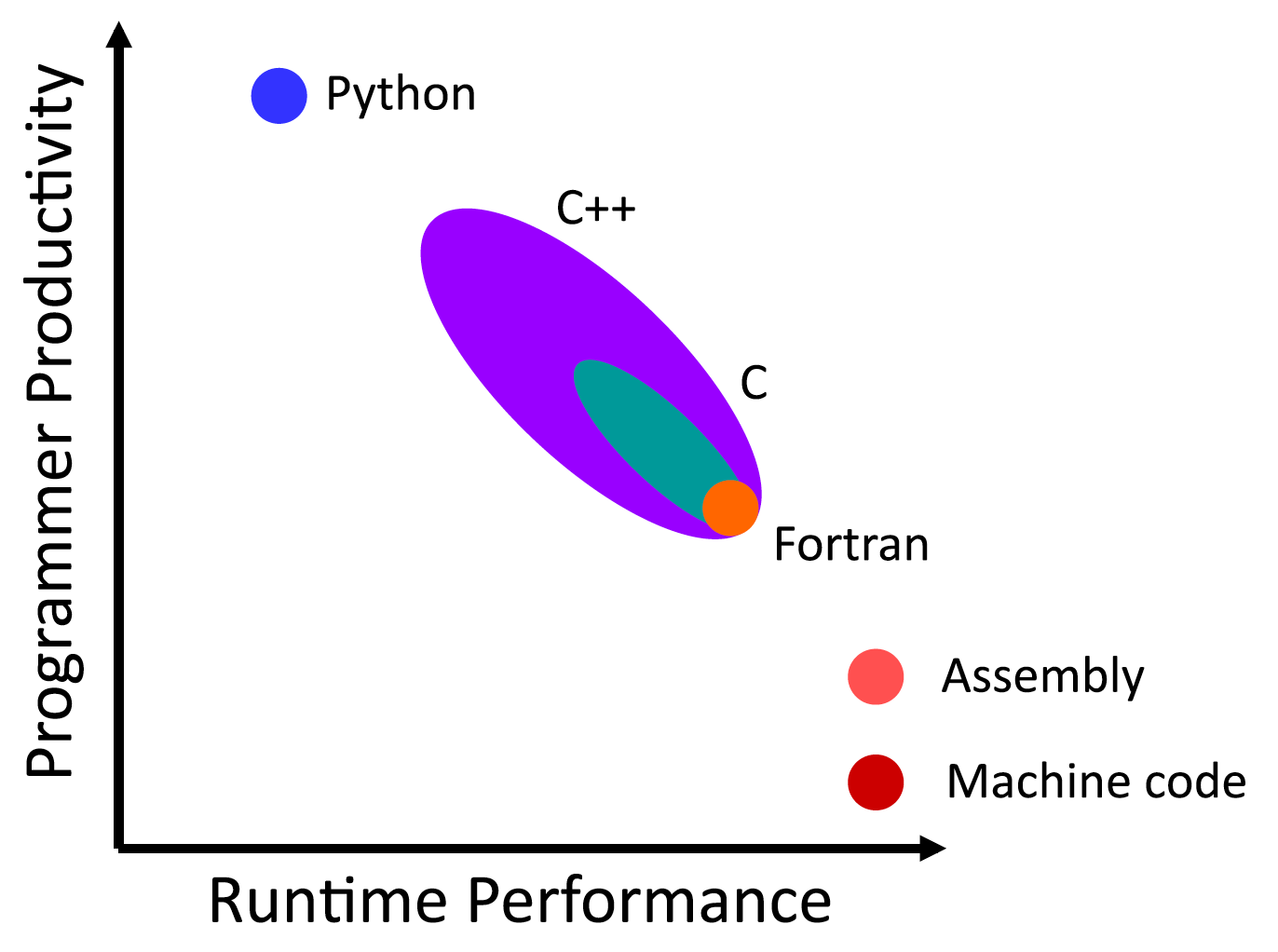}
  \caption{Qualitative graph of run-time performance vs. programmer productivity for selected programming languages \cite{grosse2012automatic}.}\label{fig:4}
  \end{subfigure}
  \caption{Code Languages Top Feature Comparison in the Case of Numerical Simulations.}
 \label{figure4}
\end{figure}

  Finally, and most importantly, although Fortran is not considered to be within the most common programming languages for general purposes among developers due to its relative simplicity and lack of abstraction, we found these disadvantages as advantages, as Fortran is just sufficient to adapt to the programming style of the 21st century, while holding the superiority in performance, in the sense of virtue which cancel the 'flaw'. Figure \ref{fig:4} compares different programming languages by demonstrating the trade-off between programmer productivity and run-time performance, and places Fortran in the exact needed position for HPC and scientific calculation purposes - almost best in terms of performances (secondary only to machine code, which is not a feasible option), and in the realm of C and C++ in terms of programmer productivity. Therefore, integrating modern design with high performance computations makes Fortran the most suitable programming language for our end.



  \subsection{ \textbf{Code Methodology:} \textit{OOPP}} 
  Working with legacy code that is programmed via functional programming has led to the following main issues:
\begin{itemize}
    \item \textbf{Scalability}: Having the data accessed by a number of functions in different locations in the code imply that this data element has to be global for these functions. Thus, an extensively used data element rises further towards the top of the dependency function tree as it has to be global to a lot of functions.
    \item \textbf{Maintainability}: Maintaining a large software by upgrading and modifying it tends to be a very tedious task, especially because usually functions depend on several other functions and data structures. Thus, even a small change in a function or a data structure might require wide changes in all the depending functions or data structures.
    \item \textbf{Readability}: new users tend to spend a large amount of time in order to understand how to use the code properly. 
\end{itemize}
Introducing Object Oriented Programming (OOP) to the code helps with the above issues \cite{stroustrup2000c++} \cite{rouson2011scientific}. Programming in OOP fashion ensures that the code is designed with a big emphasis on the locality of the data, i.e. each object accesses the data that it requires from within its class. This allows distributing the data among all objects, rather than saving the data in a centralized manner in which all objects try to access the same pieces of data simultaneously. Additionally, OOP allows placing all code related to a specific physical quantity, numerical calculations or data structures into independent classes. Thus, maintaining these classes becomes a much easier task, since the updates or modifications should be made in theses classes and only there.
Finally, we found reading and understanding an OOP program an easier task as opposed to an equivalent program that is programmed in a functional programming fashion. This is due to the fact that OOP programs encourage categorizing code into classes. To a human reader, classes and instances of them (objects) automatically give initial information about the purpose and intention of lines of a code that include them. By naming conventions and a careful categorization of code into classes, an increase in the readability of the code is achieved.

Although OOP allows a high level of abstraction which may help with scalability, maintainability and readability, it raises concerns of whether the computational performances of a code are sufficient for HPC requirements. This is due to the fact that introducing efficient parallelism may incur a violation of important OOP principles, such as object encapsulation, modularity and functional independence. We employed Object Oriented Parallel Programming (OOPP) ideas which resemble other OOP based scientific codes / platforms such as \cite{pinho2014object} \cite{calvin2002object}, in which objects are intrinsically parallel. Other works show other OOPP variants that emphasize the importance of introducing OOP into HPC applications \cite{fan2017towards} \cite{radenski1998object}. Thus keeping the benefits of OOP (without breaking too-many OOP principles), while preserving the efficiency that is required from modern scientific codes on the other hand.

The following two main principal methods of parallel computing will take part in our OOPP design: distributed processing (using MPI) and shared memory processing (using OpenMP). Integrating these parallelization methods will yield a hybrid parallelization scheme, which combines the advantages of the distributed memory (distributing the calculation over many remote processing units) and the shared memory (minimization of the communication between the processing units).

  \subsection{ \textbf{Distributed Processing:} \textit{MPI 4}}
\label{openmp}
  Today, HPC platforms make it possible for scientific software to achieve breakthroughs in many fields such as physics, biology, chemistry, etc. These breakthroughs may be explained by the following reasons: 1) the ability to model larger physical systems, 2) with greater resolution, 3) for longer periods of time, 4) and with greater fidelity to the actual physics \cite{intro-to-hpc}. Some of them are feasible directly and solely from distributed computing, which allows storing the huge amount of calculation data (that can no longer be stored on a single computing node) on several nodes, and to distribute the workload between nodes, thus possibly reducing the running time of the program by a factor of the number of processors. Message Passing Interface (MPI) is a standard for message-passing between processors. It was designed by a group of researchers from academia and industry to allow infrastructure for distributed computing for the programming languages C, C++ and Fortran. MPI is considered to be a lower level of abstraction for distributed frameworks, as opposed to Spark for example, which is automatically fault-tolerant, or Chapel, which has features like automatically reading parameters from the command line. Mostly, as the abstraction of the framework gets higher - the higher run-time penalty is yielded. In contrast, a lower-level of abstraction may penalize users for its functional simplicity. Indeed, the main problem with older versions of MPI was failure handling - both in terms of messages to the users in order to detect the failure point, as well as in the ability of a supercomputer to hold such MPI run without even a single node failure (resulting in a crash of the entire run). As a result, those topics are of main interest in MPI 4 \cite{openmpi4}, which includes extensions to better support hybrid programming model, support for fault tolerance in MPI applications and more. Moreover, the introduction of SCR (see ~\nameref{SCR}) to the code manages to allow large runs - both in terms of memory and computing power - on a supercomputer to sustain even in cases of nodes failure with a minimum cost. We use OpenMPI 4 \cite{openmpi4}, which is an Open-Source implementation of MPI 4 in \textit{BACKUS}. We stress again, that MPI 4 is best for our purposes performance-and-productivity(since MPI 4)-wise.
  
  \subsection{ \textbf{Shared-Memory Processing:} \textit{OpenMP 4.5}}
  \label{openmp}
  The shared-memory model describes multiple processing units executing in parallel a common task, e.g. an intensive \textit{for-loop} code. These processing units share the same address space and reside on the same physical machine. Thus, communication is immediate with minimal overhead. However, this may cause memory synchronization problems. OpenMP \cite{dagum1998openmp} - a directive oriented library - is one way to implement this model. A common way to use OpenMP in scientific numerical codes is to target an intensive-computational \textit{for-loop} and insert appropriate OpenMP directives onto that loop. This allows the intensive loop to be executed on several processing units (i.e. threads) simultaneously in a coordinated manner, thus dividing the workload between the working threads. Nowadays OpenMP (version 4.5)~\cite{van2017using} includes hundreds of directives, many of which were added in version 4.0 to utilize the increasing trend of heterogeneous architectures (NUMA, co-processors and accelerators). 
  
  \begin{itemize}
  \item \textbf{Working Efficiently on NUMA Architecture}: The following two directives, which were added in OpenMP 4.0, can be useful in order to utilize NUMA architectures: \textit{Thread Affinity} and \textit{Thread Placement}~\cite{eichenberger2012design}. The \textit{Thread Affinity} construct allows the user to control how OpenMP will spread the team of threads across the machine. The three main options for thread spreading are \textit{master}, \textit{close} and \textit{spread}. In the \textit{master} policy, each thread is assigned to the same 'place' (related area in the memory) as the master thread. In the \textit{close} policy, the threads are placed close to the master. In the \textit{spread} policy the threads are spread as evenly as possible. In cases of high memory/bandwidth, the user should consider to spread the threads across the machine as much as possible to maximize the throughput. Alternatively, in cases of low memory usage and many instructions, the user might want to make sure the threads are working on the same cache-line. Once the user defines the \textit{Thread Affinity} policy it is also useful to use \textit{Thread Placement}, that states on which hardware specifically (on core-level, on socket-level, or on hyper-thread level) the threads will run.
  
  \item \textbf{Utilization of Vectored Capabilities}: Another directive added in OpenMP 4.0 is the \textit{Single Instruction Multiple Data} or \textit{SIMD} construct~\cite{klemm2012extending}. The \textit{SIMD} directive allows the execution of the same operation on multiple data elements on the hardware level. Using this directive, the processing time of the operation can potentially be reduced by a factor of the vector length of the specific hardware. Alongside, a major addition included in OpenMP 4.0 is the \textit{target} construct, which allows OpenMP to support offloading code segments into co-processors (Xeon-Phi) and/or accelerators (GPGPUs), thus providing full support to heterogeneous systems which is a necessity for modern scientific codes, as their vectored capabilities enhance parallelization by orders of magnitude. The offload construct includes copying the arrays/variables and the code from the host to the accelerator, which then executes the code. Finally it copies the arrays/variables back from the accelerator to the host. This whole process is an expensive task, therefore, it is advisable to include with the offload some sort of an if-clause that checks the intensity (memory usage and number of iterations) of the task in order to determine whether the task should be offloaded or not, thus avoiding the overhead caused by offloading in some cases \cite{lustig2013reducing}.
\end{itemize}

  There are several other mainstream shared memory parallelization platforms such as OpenACC~\cite{farber2016parallel}, OpenCL~\cite{stone2010opencl} and CUDA~\cite{nvidia2010programming}. OpenACC is a directive-based library like OpenMP. However, the support for devices and compilers alike regarding OpenACC is limited and very selective while OpenMP is widely supported and used. OpenCL provides a standard interface for shared-memory model parallelism, that is suited for heterogeneous architectures like OpenMP. However, OpenCL was found in practice to be more difficult to program \cite{memeti2017benchmarking}. CUDA is an API for accelerators developed by Nvidia suited for C, C++ and Fortran. Unlike CUDA, OpenMP allows insertion of most directives to the code without changing it. Moreover, while CUDA's support for GPUs provides excellent results, its support for CPU is nonexistent. On the other hand, OpenMP can alternate between CPUs, co-processors and GPUs, and provides almost equal results on both CPUs and GPUs~\cite{memeti2017benchmarking}~\cite{gayatri2018case}~\cite{beyer2015comparing}. For these reasons and the fact that performance differences are negligible we chose OpenMP over OpenCL, OpenACC and CUDA.

  \subsection{ \textbf{Distributed Fault-Tolerance:} \textit{SCR}} \label{SCR}
  As the computation scales over many MPI processes that are distributed over several different computation nodes, the risk of software and hardware failures grows with it. This raises the need for a scalable and reliable fault tolerance system. For that purpose, we use the Scalable Checkpoint/Restart (SCR) Library \cite{moody2010design} \cite{scrGit} \cite{scrDocs}. SCR, developed by Lawrence Livermore National Laboratory (LLNL), deploys multilevel checkpointing \cite{moody2010design}. That is frequent, fast, but more volatile checkpoints to the node local RAM/Disk that may be used to recover from software faults, and less frequent, slower but persistent checkpoints to the distributed file system (DFS) that can withstand substantial software failures. SCR also adopts several different redundancy schemes such as partner and XOR/RAID between nodes. It allows fully customizable storage hierarchy, redundancy schemes and checkpoint frequency, thus optimizing the checkpoint/restart overhead and resilience to fit the user's needs. SCR capabilities also include automatic job restart, guidance for optimal checkpoint frequency, scalable bandwidth and asynchronous data transfers to the parallel file system. We found SCR best fitting for \textit{BACKUS} as it sacles well with MPI, easy to integrate and is customizable to our needs.
  
  \subsection{ \textbf{Parallel I/O Formats:} \textit{Silo + HDF5 + PDB}}

  The requirement for higher resolution of HPC simulations and likewise larger output files, has led to the need for Distributed File Systems (DFS) which makes it possible to utilize several processing units for output writing - as opposed to writing the output in a centralized manner which leads to high bandwidth due to the bottleneck that is yielded. However, creating many files (i.e. a file per each process) and operating I/O operations to it in a parallel fashion leads to failures of DFSs due to high management costs, as DFSs tend to accept small amount of very large files and distribute them over the disks, but fail to do so in cases that there are many (small) files that are distributed over many disks with very high metadata management overhead \cite{depardon2013analysis}. For that reason, several file formats (HDF5 \cite{folk2011overview}, PDB \cite{brown1995creating} \cite{pdbUserManual}) are designed to store and organize large amounts of data in a single file from many I/O channels.

  In cases where the files are designated for visualization purposes, there is a need to organize and store the data in a special file formats, as it should be rendered in a parallel visualization software. Silo format \cite{siloUserGuide} is a visualization format intended for reading and writing data into binary files through an Application Programming Interface (API). The files that Silo produces, and the information within them, can be easily shared and exchanged between independent programs running on different computing platforms. As a result, Silo API allows the development of generic tools for scientific data processing. A common tool that process Silo output files is VisIt visualization tool (see \nameref{parallelvis}).
  
  Architecturally, Silo's format is divided into two main parts: a top-level API, and an under-level read-write (I/O) interface called a driver. The Silo format supports a variety of I/O device drivers, the most common ones are the Hierarchical Data Format 5 (HDF5) and Portable DataBase (PDB).
  
  Although Silo library functions are serial, they have key properties that allow efficient and scalable parallelization with the PMPIO platform \cite{fineberg1996pmpio}. Silo's Library I/O is programmed in C, but is also partially supported by Fortran.
  
Silo supports gridless (point) meshes, structured meshes, unstructured-zoo and unstructured-arbitrary-polyhedral meshes, block structured AMR meshes, constructive solid geometry (CSG) meshes as well as piecewise-constant (e.g. zone-centered) and piecewise-linear (e.g. node-centered) variables defined on these meshes. Silo also provides a wide range of additional configurations, and thus provides full support for parallel visualization.
    
  \subsection{ \textbf{Parallel Unit Testing:} \textit{pFUnit}} 

  Automatic and flexible unit test generation for legacy HPC code is crucial \cite{hovy2016towards}. In order to test the correctness of our code we use pFUnit \cite{pfunit} - a unit testing framework for serial and MPI-parallel (with initial support for OpenMP) software written in Fortran. The framework was originally created by developers from NASA and NGC TASC \cite{rilee2014towards}. pFUnit uses modern Fortran techniques such as Object Oriented Programming and offers a light and easy-to-use platform for Fortran developers to write and run unit tests, that verify proper behavior of specified pieces of code. 
  Since the development of new scientific software is usually heavily based on scientific paradigms and/or legacy codes, that has been extensively tested (physically and numerically), the most important benefit that is achieved by isolating each piece of code into units and testing them independently is the ability to refactor the code or to upgrade system libraries at a later date, and making sure that the units still work correctly (Regression Testing) compared to the legacy code/physical phenomena. The working methodology is to write test cases for all main functionalities so that whenever a code modification causes a fault, it can be quickly identified. Thus by using unit testing, one can detect and fix the exact unit in which the failure occurred. We wish to stress again that we use pFUnit not only for its support for Fortran but also for its support for parallel programs. Additionally, unlike other unit testing frameworks for Fortran (e.g. FRUIT \cite{fruitHomepage}), we found pFUnit to be more
  expressive, as it offers a wider support for assertion directives and intrinsic testing functions, therefore it suits better for our software engineering purposes.

  \subsection{ \textbf{Parallel Visualization:} \textit{VisIt}}
  \label{parallelvis}
  Using scientific visualization tools may deepen the scientist's understanding of the physical simulations. Over the years there is a growing demand for a nano-scale resolution simulations. This demand causes both the data produced by the simulation and the operation performed in the simulation to increase which result in longer response time. This leads to a need for a fast response-time, scalabe and parallel visualization tool. We considered in \textit{BACKUS} two visualization tools: VisIt and ParaView.
   \begin{itemize}
  \item \textbf{Visit}:
  VisIt is an Open-Source, multi-platform, interactive, parallel, two and three-dimensional scientific visualization tool developed by Lawrence Livermore National Laboratory (LLNL) that supports multiple scientific data formats. Using VisIt, users can apply different mathematical expressions and operators on the visualized data and save the visualized data for post-processing actions such as debugging, interpolation, presentation, and animations.
  Among VisIt's options are contours, isosurfaces, vector fields, color-maps, probing on a specific domain and more. 
  
  \item \textbf{ParaView}:
  ParaView, developed by Los Alamos National Labratory, is also an Open-Source, multi-platform, interactive, parallel, two and three-dimensional scientific visualization tool. Unlike VisIt, Paraview supports virtual reality (VR), which may be used by the scientist in order to broaden his understanding of the physical phenomena.
  \end{itemize}
  Both VisIt and ParaView use MPI for distributing the data among different nodes in the architecture. However, VisIt provides another layer of parallelization: VisIt-OSPRay \cite{wu2018visit}. VisIt-OSPRay is an hybrid-parallel (i.e parallelization in both distributed and shared memory) rendering system in VisIt. It allows the user to fully exploit the advantageous of the heterogeneous system for better performances. This extension has the potential to achieve a speed-up of up to an order of magnitude.
  In spite of the fact that ParaView has a clear advantage in the VR domain, VR is a feature but not a necessity, and therefore we use VisIt (and VisIt-OSPRay) in \textit{BACKUS}.
  
  \subsection{ \textbf{Integrated Development Environment (IDE):} \textit{Photran}} 

  Photran \cite{photran} is an IDE for Fortran 77, 90, 95, 2003 and 2008 based on Eclipse and the CDT. The project is maintained by the University of Illinois at Urbana-Champaign and IBM. Photran includes number of sophisticated features that are designed to make it easier to write, modify, search, and maintain Fortran codes. These include content assist, which can “auto-complete” variable and function names as the user types; a declaration view, which can show the leading comments for the selected variable or procedure; Fortran Search, which allows you to find declarations and references to modules, variables, procedures, etc.; and refactoring, which changes your source code to improve its design while preserving its behavior. Photran also provides the functionality to parse a Fortran program and construct its Abstract Syntax Tree (AST) representation. The produced AST is rewritable, i.e. Photran’s API allows AST manipulation and generation of the corresponding Fortran code. Also, the constructed AST is augmented with information about the \textit{binding} of program's entities (variables, subprograms, interfaces, etc.). This might be useful for transforming the information gained by the AST into a well-optimized code by introducing parallelism to it \cite{negara2010automatic}. We chose Photran mainly since it is an Open-Source platform that allows content assist, which improves the productivity of Fortran developers.

  \subsection{ \textbf{Version Control:} \textit{Git}} 
  
  While centralized systems were the version control system of choice for nearly a decade, Git \cite{git} - a distributed version-control - has surpassed them in recent years \cite{de2009software}. Unlike SVN, Git utilizes multiple repositories: a central repository and a series of local repositories. Local repositories are exact copies of the central repository complete with the entire history of changes. The main advantages of this method are faster commits (derives from local vs. central repository work-fashion) and no single point of failure (less prone to code breaks and central repository stoppage as users work with local repositories, which are also offline) \cite{brindescu2014centralized}.

  \subsection{ \textbf{Documentation:} \textit{Doxygen}}  
  Unlike past legacy software, which used to be pre- and post-development documented, apart from the code itself, the main notion of Doxygen \cite{doxygen} is that the code documentation is \textit{dynamic} and is written \textit{within} the code body, and is generated, by demand, to a complete and live UML (Unified Modeling Language) structure of the code, as well as to documentation files. In this way, it supports Agile development and up-to-date information about the code, which is essential in building a long-lived code, as current code projects are no longer pre-designed in Waterfall methodology, thus there is no detailed code structure and documentation up-front to the development of the code. However, there is still a need for such a detailed structure in order to keep the design and development wise and agile. Doxygen also supports many programming languages - especially the low-level ones - and runs on most of operating systems.

  \subsection{ \textbf{Project Managment:} \textit{Tuleap}}
  The process of building and maintaining a code involves many tasks and requires many frequent changes and revisions of the code. This process is usually a task for a team rather than a single programmer. In order to coordinate all these tasks among the team members, one should consider the usage of an Application \textit{Lifecycle Management (ALM)} software. We used an Open-Source ALM software called \textit{Tuleap}~\cite{tuleap}. Tuleap provides a user-friendly web interface that supports the entire lifecycle of the software. It collects development tasks and inspected bugs, maintains the code using modern source-control tools (such as Git and SVN), provides a virtual environment to perform efficient code reviews, and interfaces with MediaWiki and various continuous-integration tools. Tuleap environment contains a referral mechanism that enables the user to tie between records such as MediaWiki articles, development tasks, application bugs, Git commits and so on. The ALM software supports the management process and collects the history of the development process.

  \subsection{ \textbf{User Interface:} \textit{ICE}} \label{ice}
  One of the main obstacles in a scientific research caused by the fact that in order to make minor (yet important) changes in any of the simulation software aspects, such as changing one of the simulation's parameters, the researcher has to be very familiar with the code and its structure which, in many cases, is very difficult to attain. ICE (Eclipse Integrated Computational Environment)~\cite{icepaper} provides a user-friendly interface for users to make code modifications, analyze results and re-run jobs either on a local or on a remote machine. 
  ICE improves work efficiency and shortens the overall time spent on the 'front-end engagement' of the user with the scientific software. This improvement is crucial as scientific codes tend to be very complex to handle, usually run on complicated machine architectures with minimal operating system support, and generally requires an involvement of many different software coupling until reaching the needed result~\cite{oren2017calcul}. In contrary, ICE is based on Eclipse's plugins which allows quick code modifications; Data file changes over GUI; Simple connection to remote machines; and provides visualization support (VisIt~\cite{VisIt} and ParaView~\cite{ParaView}, see \nameref{parallelvis}). ICE is built first and foremost for scientific usage, hence provides rich all-inclusive interface for update, run, analyze and visualize simulations and their results.

 \section*{Acknowledgments}
 This work was supported by the Lynn and William Frankel Center for Computer Science. Computational support was provided by the NegevHPC project \cite{negevhpc}.

\bibliographystyle{unsrt} 
\bibliography{bibliography} 

\end{document}